\documentclass[9pt,shortpaper,twoside,web]{ieeecolor}

\usepackage{generic}
\usepackage{cite}
\usepackage{amsmath,amssymb,amsfonts}
\usepackage{algorithmic}
\usepackage{graphicx}
\usepackage{textcomp}
\usepackage[export]{adjustbox}
\usepackage{subfig}
\usepackage{float}
\usepackage{hyperref}
\usepackage{multirow}
\UseRawInputEncoding

\usepackage{tikz}
\usetikzlibrary{spy}

\def\BibTeX{{\rm B\kern-.05em{\sc i\kern-.025em b}\kern-.08em
    T\kern-.1667em\lower.7ex\hbox{E}\kern-.125emX}}
\begin{document}


\title{Self-learning for weakly supervised Gleason grading of local patterns}

\author{
Julio Silva-Rodríguez, Adrián Colomer, Jose Dolz and Valery Naranjo
\thanks{This work was supported by the Spanish Ministry of Economy and Competitiveness through projects DPI2016-77869 and PID2019-105142RB-C21.}
\thanks{Julio Silva-Rodríguez is with the Institute of Transport and Territory, Universitat Polit\`ecnica de Val\`encia, Valencia, Spain (e-mail: jjsilva@upv.es).}
\thanks{Adrián Colomer and Valery Naranjo are with the Institute of Research and Innovation in Bioengineering, Universitat Polit\`ecnica de Val\`encia, Valencia, Spain (e-mail: {adcogra,vnaranjo}@i3b.upv.es).}
\thanks{Jose Dolz is with the ETS Montreal, Montréal, QC, Canada (e-mail: jose.dolz@etsmtl.ca).}
\thanks{© 2021 IEEE.  Personal use of this material is permitted.  Permission from IEEE must be obtained for all other uses, in any current or future media, including reprinting/republishing this material for advertising or promotional purposes, creating new collective works, for resale or redistribution to servers or lists, or reuse of any copyrighted component of this work in other works.}
}

\maketitle

\begin{abstract}

Prostate cancer is one of the main diseases affecting men worldwide. The gold standard for diagnosis and prognosis is the Gleason grading system. In this process, pathologists manually analyze prostate histology slides under microscope, in a high time-consuming and subjective task. In the last years, computer-aided-diagnosis (CAD) systems have emerged as a promising tool that could support pathologists in the daily clinical practice. Nevertheless, these systems are usually trained using tedious and prone-to-error pixel-level annotations of Gleason grades in the tissue. To alleviate the need of manual pixel-wise labeling, just a handful of works have been presented in the literature. Furthermore, despite the promising results achieved on global scoring the location of cancerous patterns in the tissue is only qualitatively addressed. These heatmaps of tumor regions, however, are crucial to the reliability of CAD systems as they provide explainability to the system's output and give confidence to pathologists that the model is focusing on medical relevant features. Motivated by this, we propose a novel weakly-supervised deep-learning model, based on self-learning CNNs, that leverages only the global Gleason score of gigapixel whole slide images during training to accurately perform both, grading of patch-level patterns and biopsy-level scoring. To evaluate the performance of the proposed method, we perform extensive experiments on three different external datasets for the patch-level Gleason grading, and on two different test sets for global Grade Group prediction. We empirically demonstrate that our approach outperforms its supervised counterpart on patch-level Gleason grading by a large margin, as well as state-of-the-art methods on global biopsy-level scoring. Particularly, the proposed model brings an average improvement on the Cohen's quadratic kappa ($\kappa$) score of nearly $18\%$ compared to full-supervision for the patch-level Gleason grading task. This suggests that the absence of the annotator's bias in our approach and the capability of using large weakly labeled datasets during training leads to higher performing and more robust models. Furthermore, raw features obtained from the patch-level classifier showed to generalize better than previous approaches in the literature to the subjective global biopsy-level scoring.

\end{abstract}

\begin{IEEEkeywords}
Gleason grading, Prostate cancer, Self-learning, Weakly supervised, Whole slide images.
\end{IEEEkeywords}

\section{Introduction}
\label{sec:introduction}



\begin{figure}[htb]
\captionsetup[subfloat]{farskip=1pt,captionskip=0.8pt}
    \centering
      \subfloat[\label{fig1a}]{\includegraphics[width=.235\linewidth, frame]{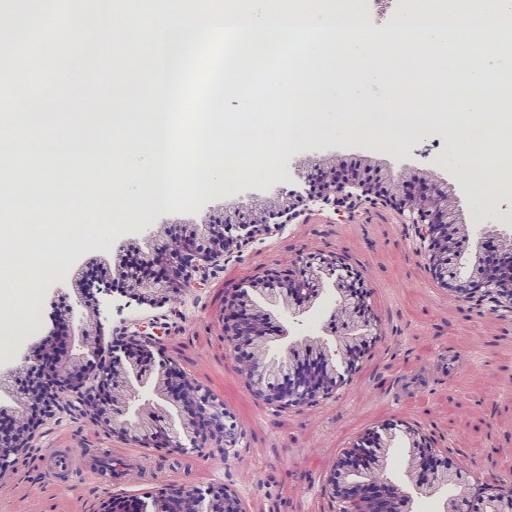}}
      \hspace*{\fill}
      \subfloat[\label{fig1b}]{\includegraphics[width=.235\linewidth, frame]{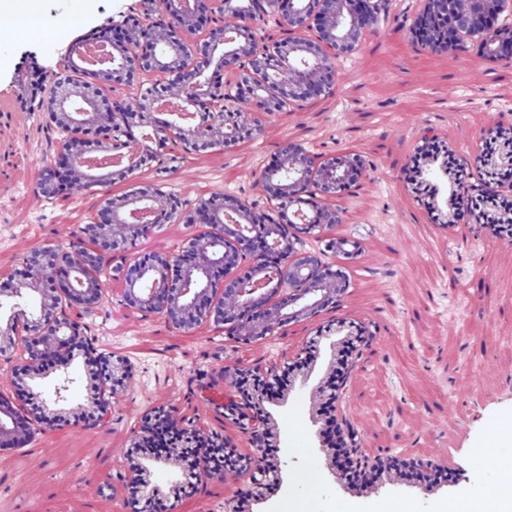}}
      \hspace*{\fill}
      \subfloat[\label{fig1c}]{\includegraphics[width=.235\linewidth, frame]{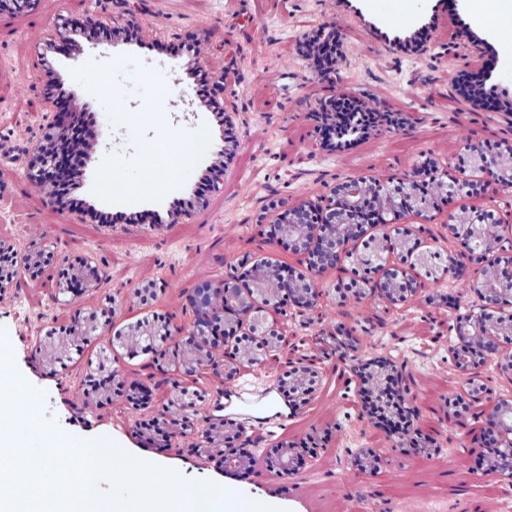}}
      \hspace*{\fill}
      \subfloat[\label{fig1d}]{\includegraphics[width=.235\linewidth, frame]{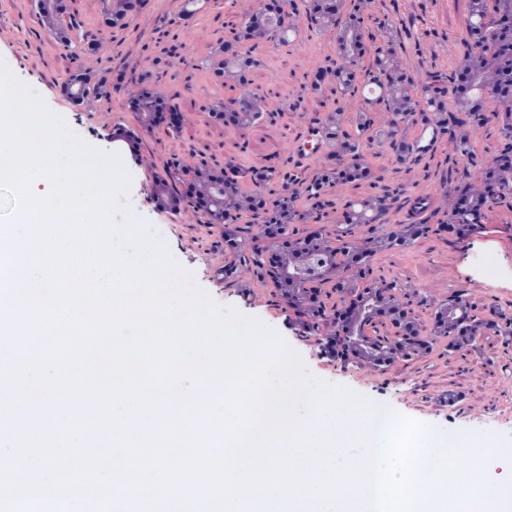}}
    \caption{Histology regions of prostate biopsies. (a): region containing benign glands, (b): region containing GG3 glandular structures, (c): region containing GG4 patterns, (d): region containing GG5 patterns. GG: Gleason grade.}
    \label{fig1}
\end{figure}

Prostate cancer is one of the major diseases affecting men worldwide. It accounts for $14.5\%$ of all cancers in men \cite{wcrf} and, according to the World Health Organization, its yearly incidence will increase to 1.8 million cases this decade \cite{who}. The Gleason grading system \cite{gleason} is the main tool for its diagnosis and prognosis. This system describes different stages of cancer based on the morphology and distribution of glands in prostate biopsies. Specifically, the Gleason grades (GG) observable in histology samples range from $3$ (GG3) to $5$ (GG5). Fig. \ref{fig1} shows representative patterns of each grade.

In order to make a diagnosis of prostate cancer, pathologists extract small portions of tissue, which are laminated and stained using Hematoxylin and Eosin. Then, the slides are carefully analyzed under the microscope to grade local glandular patterns according to the Gleason grading system. Finally, the two most prominent grades in terms of proportion and severity are used to obtain a global Gleason score as prognosis marker. For instance, the Gleason score $3+5=8$ would be assigned to a sample in which the main cancerous Gleason grade is GG3 and the second is GG5. Recently, after the $2014$ conference of the International Society of Urological Pathology, a new grading system referred to as Grade Group \cite{Epstein2015ASystem} has been adopted. This systems takes into account the  different prognosis between patients with Gleason score $3+4=7$ and $4+3=7$, including them to different groups (Grade Group $2$ and $3$, respectively). The whole diagnostic process is highly time-consuming, and is characterized by a large variability among pathologists \cite{Burchardt2008InterobserverMicroarrays}. These limitations have motivated the development of automatic tools to analyze whole slide images in recent years.

Computer-aided diagnosis (CAD) systems based on computer vision algorithms are able to support pathologists in the daily analysis of prostate biopsies. However, the development of these applications is limited, mainly due to the high data-demanding nature of deep learning algorithms, the large size of digitized biopsies (known as whole slide images (WSIs)) and the difficulty in obtaining pixel-level labeled histology images \cite{Komura2018MachineAnalysis}. Current CAD systems are usually developed to classify local cancerous regions, which are finally combined into a global score. In the case of prostate cancer, this requires manual annotation using multi-resolution graphical user interfaces to accurately delimit the cancerous structures using the Gleason grading system. This is a laborious process, prone to error due to pathologists discouragement, which could incorporate the annotator's bias for certain patterns. Moreover, heterogeneous epithelial cancer such as prostate cancer requires a large number of samples to cover the wide range of possible patterns, which is difficult to reach on annotated datasets.    


The limitations of using large annotated datasets encourages the development of weakly-supervised methods able to leverage global labels --easily accessible from the clinical record via the Gleason score-- during the training process. Nevertheless, the literature on employing the global Gleason score to develop CAD systems for prostate biopsy grading remains scarce. The main limitation of the proposed approaches is that they focus on the global-level scoring, while the challenge of local grading cancerous patterns is only qualitatively validated, or simply not addressed. It is noteworthy to mention that the classification of local Gleason grades in prostate biopsies is the basis of an explainable prostate CAD system. The resultant heatmaps support the biopsy-level scoring provided by the system, and they demonstrate that the model relies on relevant medical markers. Thus, the accuracy of the proposed methods in this task must be validated, in order to provide confidence to pathologists in the daily-use of CAD systems. 


In this work, we propose a novel weakly-supervised learning strategy to perform both, the global scoring of biopsies and the local grading of cancerous structures in the tissue, where learning is driven only by the global Gleason score. To the best of our knowledge, this is the first attempt to accurately grade local cancerous patterns in prostate whole slide images using biopsy-level labels during training. In the following lines, we summarize the main contributions of this paper. First, we propose an end-to-end CNN architecture, based on patch-level inference aggregation, that is able to detect high-confidence cancerous instances in a weakly-supervised multiple-instance learning (MIL) scenario. Then, we propose a self-learning framework that converts the MIL dataset into a pseudo-supervised task, employing the patches predicted by the previous model and a subsequent post-processing label refinement. We empirically demonstrate that weakly-supervised models trained on large datasets are able to generalize better on the patch-level Gleason grading task than supervised models trained in smaller databases with pixel-level annotations. Finally, we predict the global biopsy-level score based on the aggregation of local features by using the models trained for the patch-level Gleason grading.

\section{Related Work}
\label{sec:relWork}

\subsection{Self-Learning}
\label{ssec:rw1}

In the context of this work, we refer to self-learning (a.k.a. self-paced learning or self-training) as the training procedure introduced in \cite{Scudder1965ProbabilityMachines}, which aims to use the knowledge of a firstly trained model (usually called teacher) into a second model (known as student). Interest in this technique has grown in recent years due to the promising results obtained in semi- and weakly-supervised learning scenarios. For instance, in semi-supervised learning approaches, the teacher is used to obtain pseudo-labels from non-labeled data, after it has been trained on annotated examples \cite{Xie2020Self-trainingClassification,Yalniz2019Billion-scaleClassification, Veit2017LearningSupervision}. Afterwards, the student model is trained by integrating the pseudo-labels in the augmented training dataset. To train more robust students, which are also consistent with the teacher, \cite{Xie2020Self-trainingClassification} introduces noise to the samples, as well as model noise, while \cite{Yalniz2019Billion-scaleClassification} selects the top–K images based on the
corresponding classification scores by the teacher. These works also exploit knowledge distillation by transferring the teacher knowledge to either larger \cite{Xie2020Self-trainingClassification} or smaller \cite{Yalniz2019Billion-scaleClassification} students. In the context of weakly-supervised object localization, several works employ a teacher model to select regions of interest from the image to train student model in a simplified dataset \cite{Bazzani2016Self-taughtNetworks, Sangineto2019SelfDetection, Jie2017DeepLocalization}. 
We want to emphasize that the techniques presented here differ from the similarly so-called self-learning methods, which pre-train networks on pretext tasks where both the inputs and labels are derived from an unlabeled dataset \cite{Chen2020BigLearners,patacchiola2020self}\footnote{The terminology used in this work is proposed by analogy to the previous work on self-learning applied to semi-supervised learning in \cite{Xie2020Self-trainingClassification}.}. Even though both techniques aim at leveraging unlabeled, or weakly unlabeled data, they present fundamental and methodological differences.

Inspired by these previous works, we adopt a self-learning strategy to accurately classify instances using image-level labels from WSIs. Nevertheless, our work differs from these in that our strategy: (1) trains a teacher model on global image labels, (2) uses the teacher predictions to generate pseudo-labels on unlabeled instances at patch-level, and (3) trains a student model, of the same complexity, on the pseudo-labels generated by the teacher.

\subsection{Multiple Instance Learning}
 
The multiple instance learning (MIL) paradigm falls under the umbrella of weakly supervised learning. In this setting, the training instances are grouped in sets, referred to as bags, $X$, where only the label for an entire bag, $Y$, is known. Thus, a bag is considered positive for certain class if at least one instance is positive such that:

\begin{equation}
    \label{eq:MIL}
       Y_k=
       \begin{cases} 
        1, & \text{if}\ \exists \; x : y_k = 1\\ 
        0, & \text{otherwise}
        \end{cases}
\end{equation}

With the advent of deep learning, recent efforts on this field have focused towards training a feature extractor under the MIL framework using CNNs. Then, a bag-level representation is obtained by aggregation of either the instance-level features (\textit{embedding-based}) or predictions (\textit{instance-based}). Typical aggregation functions include the maximum \cite{Oquab2012IsFree}, average, and log-sum-exponential \cite{Pinheiro2015FromNetworks} pooling, more advanced min-max mechanisms recently proposed, such as WILDCAT \cite{Durand2017WILDCAT:Segmentation}, and trainable functions such as AttentionMIL \cite{Ilse2018Attention-basedLearning}. Most recent works on MIL adress the problem of weakly-supervised segmentation. In this scenario, embedding-based methods are employed to obtain pixel-level predictions via gradient methods (e.g., grad-CAMs \cite{Selvaraju2020Grad-CAM:Localization}), which are later refined via self-training iterative strategies \cite{Lee2019Ficklenet:Inference, Wang2020Self-SupervisedSegmentation}. Nevertheless, it is noteworthy to highlight that weakly-supervised segmentation works with co-dependent instances (pixels), which are merged on combined features in the CNN. Thus, the generalization of these methods to MIL scenarios which use images as instances is not straightforward.

Nonetheless, despite the wide adoption of MIL in computer vision, its use in prostate histology images still remains scarce. There have been only few attempts to resort to the MIL paradigm in this scenario, which are detailed on the following section. 

\subsection{Gleason Grading in Prostate Histology Images}
\label{ssec:rw2}

A reliable Computer-Aided Diagnosis system in prostate cancer using histology biopsies aim two main tasks: the global scoring of slides and the quantification and localization of cancerous tissue, both using the Gleason grading system. Due to the large dimension of WSIs and the computational limitations of CNNs, the basis of these systems is the use of small patches extracted from the slide. The proposed methods in the literature can be divided into two categories: bottom-up frameworks, which perform a patch-level classification of Gleason grades using pathologists annotation, and top-down methods, which perform a pseudo-labeling of the patches based on the global Gleason score of the sample. 

First works in this field focused on bottom-up frameworks. They usually fine-tune well-known CNN architectures in a supervised patch-level classification \cite{Nir2018AutomaticExperts,Nir2019ComparisonImages,Arvaniti2018AutomatedLearning,Silva-rodriguez2020GoingDetection}. Note that these methods require pixel-level expert annotations to obtain the ground truth. Recently, different approaches have been proposed in the literature to overcome the need of pixel-level annotations of Gleason grades. These methods are based in a top-down strategy, where global labels (easily accessible from the patient clinical record) are assigned to local regions of interest. In this way, a weakly-supervised patch-level classification model is trained using the pseudo-labels obtained from global images. In this vein, Campanella et al. \cite{Campanella2019Clinical-gradeImages}, under the MIL formulation, assigned the global label (cancerous against non-cancerous) to all the patches of the slide, resulting in a considerable amount of noisy labels. In \cite{JimenezdelToro2017ConvolutionalScore} and \cite{Otalora2019StainingPathology}, color-based filtering was employed to select only patches with high presence of nuclei in cancerous slides, and Ström et al. \cite{Strom2020ArtificialStudy} followed a similar strategy using Laplacian filters. Bulten et al. \cite{Bulten2020AutomatedStudy} proposed a semi-supervised pipeline and discarded patches that presented low amount of epithelial tissue. In that work, glandular tissue was previously segmented using an UNet trained using pixel-level annotations. Although these works provided promising results for the global biopsy-level scoring, the patch-level classification was not quantitatively validated. Few works only performed a qualitative evaluation of the produced heatmaps. In these methods, the localization of Gleason grades in the tissue could be affected by the assumptions made to obtain the patch-level pseudo-labels. Accurate localization of Gleason grades in the tissue is a major task that CAD systems should address. The produced heatmaps provide explainability to the system, and ensures that the output is based on medical factors to entrust pathologists in their daily use. Contrary to these works, we propose a teacher model based on instance-level MIL to infer patch-level Gleason grades from bag (biopsy)-level Gleason scores.

Regarding the global scoring of biopsies, the main approach is based on aggregating the patch-level predictions of Gleason grades via the percentage of each grade in the tissue. Particularly, the different models proposed to predict the global Gleason score or ISUP group include: threshold strategies \cite{Arvaniti2018AutomatedLearning,Bulten2020AutomatedStudy}, a k-nearest neighbor model \cite{Nagpal2019DevelopmentCancer}, a multilayer perceptron \cite{Silva-rodriguez2020GoingDetection}
 or random forests \cite{Strom2020ArtificialStudy}.


\section{Methods}
\label{sec:methods}

\begin{figure*}[h!]
    \begin{center}
    \includegraphics[width=.90\textwidth]{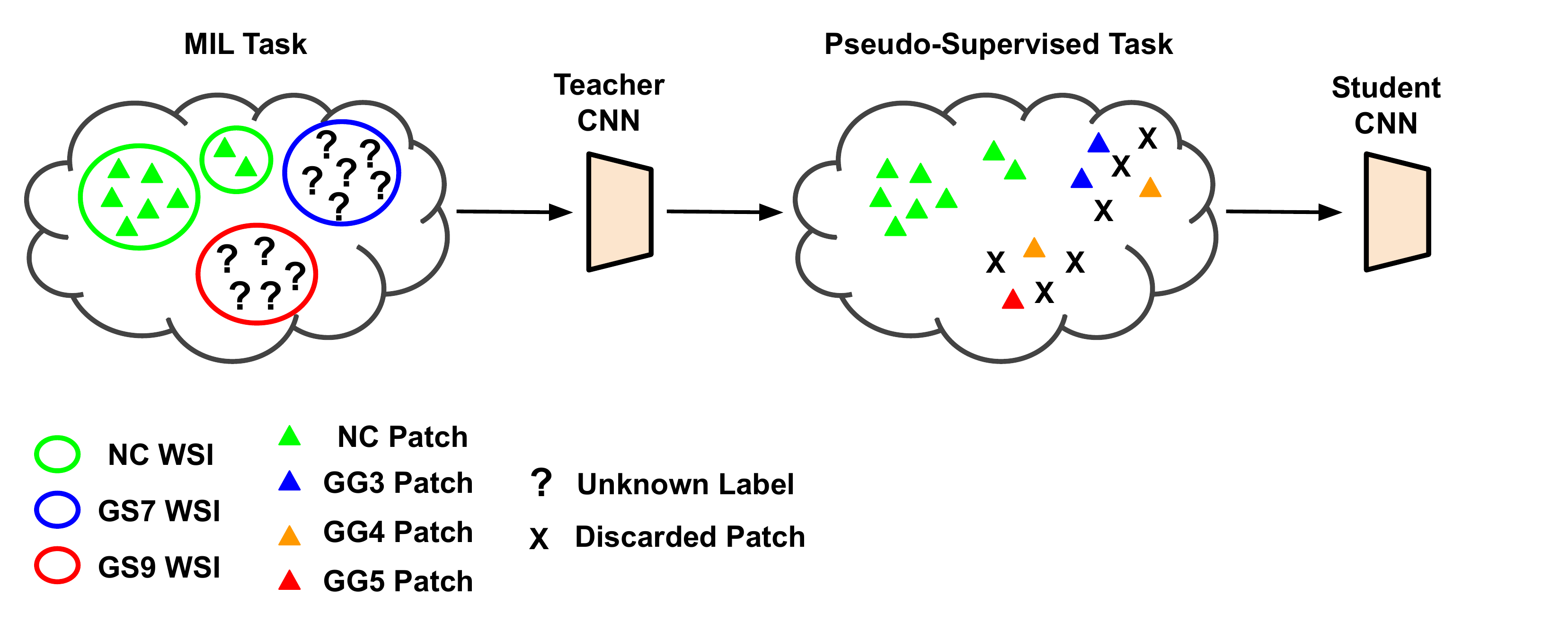}
    \caption{Self-learning CNNs pipeline for weakly supervised Gleason grading of local cancerous patterns in whole slide images. MIL: multiple-instance learning; WSI: whole slide image; NC: non-cancerous; GS: Gleason score; GG: Gleason grade.}
    \label{figselfLear}
    \end{center}
\end{figure*}

The methodological core of the proposed approach is a self-supervised CNN classifier able to grade prostate histology patches using only the biopsy-level Gleason score during training. The proposed workflow, which is composed by a teacher ($\theta^t$) and a student ($\theta^s$) model, is presented in Fig. \ref{figselfLear}. The first model, i.e., ($\theta^t$),  classifies high-confidence patches under a noisy multiple instance learning (MIL) paradigm. In this context, a prostate biopsy is considered as a bag $X_{b}$ containing instances $x_{b,i}$, and the goal is to predict the instance-level labels $y_{b,i}$ when only the biopsy-level labels $Y_{b}$ are known. $Y_{b}$ are obtained using the primary and secondary Gleason grades of the biopsy indicated in the Gleason score. 
Concretely, the non-cancerous (NC), Gleason grade (GG) $3$, $4$ and $5$ classes are included in $Y_{b}$ as a multi-label one-hot-encoding ground truth. Then, during the second step, the student model ($\theta^s$) resorts to the instance-level pseudo-labels predicted by the teacher model for training on a pseudo-supervised dataset. The details of these steps are given below.

\subsection{Teacher Model}
\label{ssec:meth1}

The teacher model aims to grade high-confidence patches using biopsy-level labels for learning. Formally, let us denote each individual bag as ${X}^{t}_{b} = \{x_{b,1},...,x_{b,I}\}$, where $x_{b,i}$ is the i-\textit{th} instance and $I$ denotes the total number of patches, i.e., instances, in the slide. Hence, the objective becomes to predict the global Gleason grade ($\hat{Y}^{t}_{b}$) from the instances ($x_{b,i}$), which can be defined as follows:

\begin{equation}
\hat{Y}^{t}_{b} = f(\{x_{b,1},...,x_{b,i},...,x_{b,I}\},\theta^t)
\label{eq:1}
\end{equation}

\noindent where $\theta$ denotes the teacher model weights.

To accomplish the inference of instance-level predictions, the learning process is based on the aggregation of patch-level predictions, i.e., $\hat{y}^t_{b,i}$. Thus, for each instance ${x}_{b,i}$ in the bag, the teacher model predicts the Gleason grade as follows:

\begin{equation}
\label{eq:tm2}
\hat{y}^t_{b,i} = f(x_{b,i},\theta^t)
\end{equation}

Then, we employ an aggregation function $p(\cdot)$ to resume all the instance-level predictions into one representative value that serves as global-level inference. Following this, eq (\ref{eq:1}) can be rewritten as:

\begin{equation}
\label{eq:tm3}
\hat{Y}^{t}_{b} = p(\{\hat{y}^t_{b,1},...,\hat{y}^t_{b,i},...,\hat{y}^t_{b,I}\})
\end{equation}

In the context of this work we employ pooling as aggregation function. It is important to mention that the pooling function should be robust to the MIL characteristics. A bag-level class could be positive if just one of the instances is positive for that class. For instance, the use of average pooling would diminish the global cancerous classes activation if the slide contains a large number of non-cancerous patches. Inspired by the properties observed in the max-pooling operation in weakly-supervised segmentation tasks \cite{Oquab2012IsFree}, we propose the use of a slide-level max-pooling. Using this operation, the global probability per class is the maximum of the patch-level inferences. This architecture ensures the classification only of high-confidence instances, since gradients in the network are only back-propagated on the instance with largest entropy. 

\begin{figure}[H]
    \begin{center}
    \includegraphics[width=.5\textwidth]{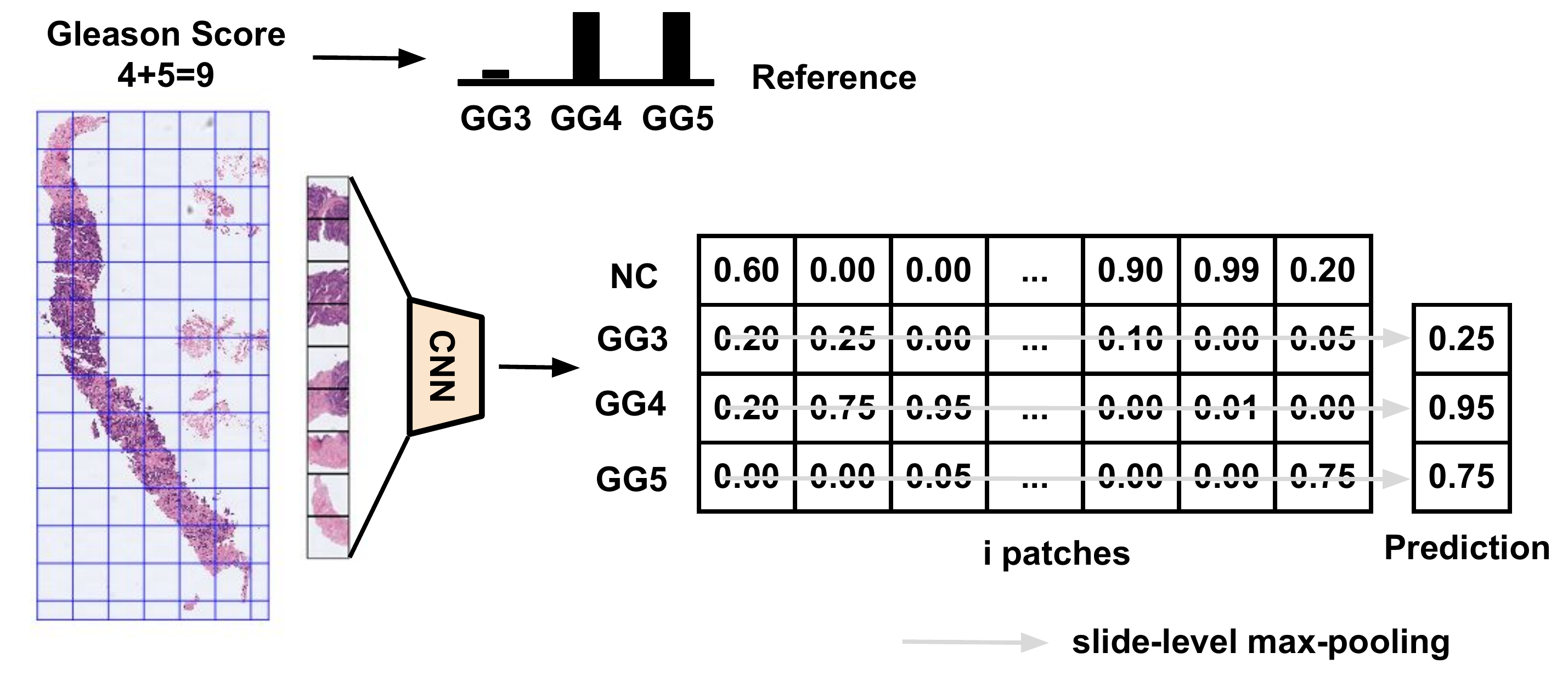}
    \caption{Teacher CNN for the prediction of local Gleason grades in a multiple-instance learning framework. GG: Gleason grade.}
   \label{figteacherModel}
    \end{center}
\end{figure}

Finally, multi-label binary-cross entropy loss is used during training for gradient estimation. Concretely, the loss is obtained using only the cancerous grades under the assumption that all slides could contain non-cancerous regions, but only cancerous slides contain patches with Gleason patterns. A summary of the Teacher model training is illustrated in Fig. \ref{figteacherModel}. 

\subsection{Student Model}
\label{ssec:meth2}

The student model aims to perform a patch-level Gleason grade prediction based on a pseudo-supervised data set of images, using the teacher model predictions as pseudo-labels. First, all instances from the dataset are predicted using the teacher model. Then, a label refinement process is carried out based on patch-level teacher model predictions ($\hat{y}^{t}_{b,i}$) and the known global slide-level labels ($Y_{b}$). During this process, labels are modified and patches are discarded under the following premises:

\begin{equation}
    \label{eq:42}
       \hat{y}{^t}_{b,i}=
       \begin{cases} 
        NC, & \text{if}\ G \not\subset Y_{b} \\ 
        G, & \text{if}\ G \subset \hat{y}{^t}_{b,i} \land G \subset Y_{b} \\
        \text{Discarded}, & \text{Otherwise}
        \end{cases}
\end{equation}
\noindent where $NC$ denotes the non-cancerous class, $G$ an undefined Gleason grade, and $\hat{y}^{t}_{b,i}$ is the hard one-hot encoding of the Teacher model prediction for certain patch $i$, belonging to the slide $b$. By doing this, only the patches classified as certain Gleason grade that belong to a slide actually containing that grade are kept for the subsequent learning of the student model. Regarding the non-cancerous patches, these are obtained only from known benign slides. This label refinement post-processing, together with the simplification of the problem from a MIL to a pseudo-supervised framework, allows the student model to better learning feature representations of the patches. Finally, the student model, which has the same architecture as the teacher, is trained minimizing the categorical-cross entropy between predictions and pseudo-labels. To account for class imbalance, class-specific weights are integrated into the loss function:

\begin{equation}
\label{eq:loss}
L(\hat{y}^s_{b,i},\hat{y}^t_{b,i}) = - \frac{1}{C}\sum_{c=1}^{C}w_{c}(\hat{y}^t_{b,i,c}log(\hat{y}^s_{b,i,c}))
\end{equation}

\noindent where $C$ is the total number of classes (i.e. non-cancerous, Gleason grade $3$, $4$ and $5$), $w_{c}= (C \times N) / N_{c}$ is the weight corresponding to each class, being $N$ the total number of images and $N_{c}$ the number of images belonging to class $c$.


\subsection{Biopsy-Level Gleason Scoring}
\label{ssec:meth3}

Once the Gleason grades are located in the tissue, we propose to use the specialized features extracted by student model to predict the global Gleason score. Thus, if we denote $z{^s}_{b,i}$ the features extracted by the student model for each patch, the slide-level feature representation is obtained by global averaging the instance-level features as follows:

\begin{equation}
\label{eq:globalscoring}
z^{s}_{b} = \frac{1}{I} \sum_{i} z^{s}_{b,i}
\end{equation}

Then, two different models are used to predict both the global Gleason score and Grade Group. First, a simple multi-layer perceptron (MLP) composed of one hidden layer with 64 neurons followed by a ReLU activation is used to predict the one-hot-encoding of the global labels. In this case, soft-max activation is used in the output layer and the weights are optimized using the categorical cross-entropy loss. Regarding the second model, k-Nearest Neighbors (kNN) is employed to compare the generalization capability of neural networks and non-parametric models for this task.

\section{Experiments and Results}
\label{sec:experiments}

In this section, we describe the experiments carried out to validate the proposed method. First, the different datasets and metrics used are presented in Sections \ref{sec:dataset} and \ref{sec:metrics}, respectively. The implementation of our self-learning strategy is detailed in Section \ref{sec:implementation}. To evaluate the benefits of the proposed approach compared to previous works, baseline and state-of-the art methods are used for comparison on the same datasets. These methods are described in Section \ref{sec:baseline}. Finally, results obtained for both tasks, i.e., grading of local patterns and global scoring of biopsies are presented in Section \ref{ssec:exp2} and \ref{ssec:exp3}, respectively.

\subsection{Experimental setting}
\subsubsection{Dataset}
\label{sec:dataset}

The experiments described in this paper are conducted using several public datasets, which are well known in the prostate cancer histology literature. Concretely, two different datasets of prostate WSIs are used to validate the global biopsy-level methodology, while three databases with pixel-level annotations are used to test the local Gleason grading capability of the proposed methods.

Regarding the datasets used to validate the biopsy-level classifier performance, the recently released dataset from the MICCAI 2020 PANDA challenge \cite{wouter_bulten_2020_3715938} is used to evaluate the proposed algorithms. This dataset consists of $10,415$ prostate WSIs whose primary and secondary Gleason grades have been labeled by expert pathologists. The gigapixel images were resampled to $10\times$ resolution, and randomly clustered into three groups for training, validating and testing. Further, the external SICAP\footnote{SICAPv2 dataset is accessible at: http://dx.doi.org/10.17632/9xxm58dvs3.2.} database presented in \cite{Silva-rodriguez2020GoingDetection} is used for testing. This dataset consists of 155 prostate WSIs with both global primary and secondary Gleason grades annotated by expert pathologists. The obtained splits for the PANDA dataset, as well as the Gleason score distribution across both datasets are presented in Table \ref{tab:dataWSI}.

\begin{table}[htb]

    \centering
    \caption{Datasets of prostate biopsies used. Whole slide images partition and Gleason scores (GS) distribution.}
    \label{tab:dataWSI}

    \begin{tabular}{|l|c|c|c|c|c|c|}
    \hline
    \multicolumn{1}{|c|}{\textbf{Partition}} & \textbf{NC} & \textbf{GS6} & \textbf{GS7} &  \textbf{GS8} &  \textbf{GS9} &  \textbf{GS10}\\
    \hline
    \hline
    \multicolumn{7}{c}{PANDA} \\
    \hline
    \hline
    Train & $2,297$ & $2,122$ & $2,075$ & $1,002$ & $874$ & $99$\\
    \hline
    Validation & $98$ & $89$ & $85$ & $42$ & $33$ & $6$\\
    \hline
    Test & $497$ & $455$ & $425$ & $205$ & $190$ & $22$\\
    \hline
    Total & $2,892$ & $2,666$ & $2,585$ & $1,249$ & $1,097$ & $127$\\
    \hline
    \hline
    \multicolumn{7}{c}{SICAP} \\
    \hline
    \hline
    Test & $36$ & $14$ & $45$ & $18$ & $35$ & $7$\\
    \hline
    \end{tabular}

\end{table}

In order to validate the capability of the proposed methods to grade local cancerous patterns, three different external datasets containing pixel-level annotations of Gleason grades are used. Concretely, the test cohort from SICAP dataset and the ARVANITI \cite{Arvaniti2018AutomatedLearning} and GERTYCH \cite{Gertych2015MachineProstatectomies} databases\footnote{ARVANITI and GERTYCH datasets were obtained upon request of corresponding authors of \cite{Arvaniti2018AutomatedLearning} and \cite{Gertych2015MachineProstatectomies}, accordingly.} are used. For these sets, patches of size $512^2$ pixels are extracted at $10\times$ resolution. This choice is motivated by prior literature, which determined this configuration as the most optimum for the binary cancer vs. no cancerous supervised classification task \cite{Esteban2019AProcesses}. Furthermore, the main study on supervised learning used for comparison, i.e., \cite{Silva-rodriguez2020GoingDetection}, employs the same patch size, which makes direct comparison easier. Even though this image size might be considered large, benign fusiform or dilated glands, and cribriform GG4 structures may come to have sizes in this range, and smaller patch size could impede the visualization of complete glandular structures. For SICAP dataset, the label was assigned by majority voting of the pixel-level annotations, and for ARVANITI and GERTYCH datasets, only patches containing one Gleason grade were used. The non-cancerous class is assigned to patches containing only benign annotations. The data source and the number of patches from each dataset, as well as the Gleason grade distribution are presented in Table \ref{tab:dataPatches}.

\begin{table}[htb]

    \centering
    \caption{Datasets with patch-level Gleason grade annotations used for testing. Distribution of the patches among non-cancerous (NC) and the different Gleason grades (GG).}
    \label{tab:dataPatches}

    \begin{tabular}{|l|l|c|c|c|c|}
    \hline
    \multicolumn{1}{|c|}{\textbf{Database}} & \multicolumn{1}{|c|}{\textbf{Source}} & \textbf{NC} & \textbf{GG3} & \textbf{GG4} &  \textbf{GG5}\\
    \hline
    
    SICAP & Biopsies & $644$ & $393$ & $853$ & $232$ \\
    \hline
    ARVANITI & Tissue Micro-Arrays & $115$ & $274$ & $210$ & $104$  \\
    \hline
    GERTYCH & Prostatectomies & $32$ & $95$ & $216$ & $70$ \\
    \hline
    \end{tabular}

\end{table}

The three external databases were normalized to homogenize the color distribution to the PANDA database. More concretely, the method presented in \cite{Vahadane2015Structure-preservedImages} was used after applying a channel-wise histogram matching to the images from the external databases to a PANDA reference image. This image was selected by the pathologists involved in this work based on its structural and stain properties.

\subsubsection{Metrics}
\label{sec:metrics}

In order to evaluate the different approaches, we resort to accuracy (ACC), f1-score (F1S) per class and its average, and Cohen's quadratic kappa ($\kappa$). The last metric, $\kappa$, is the main figure of merit typically used in prostate Gleason grading. It takes into account that the Gleason system consists on a set of ordered classes, and errors between adjacent classes should be less penalized. In addition, precision and sensitivity are obtained for the non-cancerous class for better understanding of the Teacher-Student pair behavior.

\subsubsection{Implementation Details}
\label{sec:implementation}

The patch-level classification of Gleason grades was obtained using the self-learning pipeline detailed in Sections \ref{ssec:meth1} and \ref{ssec:meth2}. The teacher model takes tiles of size $256\times256$ as input, and uses VGG16 architecture as backbone with weights pre-trained on Imagenet for the feature extraction stage. Then, a global-average pooling and a dense layer with soft-max activation is used as top model. The proposed Teacher model is based on the aggregation of patch-level outputs (Gleason grade predictions). Due to the variable amount of instances in a biopsy $I$, the architecture can not be trained using a mini-batch strategy. Thus, the learning process was carried out using a batch size of $1$ slide. The number of patches per slide varies from $40$ up to $300$ in the training set. In order to avoid computational limitations, up to $200$ random patches were used in each iteration. During the learning stage, the teacher model was trained during $30$ epochs by using SGD optimizer. The learning rate ($\eta$) was initialized to $1\cdot10^{-2}$, whose value is decreased by $10$ after half the iterations. Then, an exponential decay was applied during the last $5$ epochs to stabilize the weights such that $\eta = 1\cdot10^{-3}\cdot e^{-0.1\cdot t}$, where $t$ is the epoch number. The student model was trained following the same procedure than the teacher model, i.e., same number of epochs, optimizer, and learning rate schedule.

The global scoring of biopsies was carried out using the features extracted by the student model as detailed in Section \ref{ssec:meth3}. For the multi-layer perceptron model (MLP), Adam was applied as optimizer, using a learning rate of $1\cdot10^{-2}$ during $20$ epochs. Also, a k-Nearest Neighbors (kNN) model was fitted using a $k = 20$, optimized on the validation set.

The proposed methods were implemented in Python $3.6$ using Pytorch. The scripts to reproduce the results reported in this work are publicly available on (\url{https://github.com/cvblab/self_learning_wsi_prostate}).

\subsubsection{Baseline Methods}
\label{sec:baseline}

In order to compare our proposed approach with previous state-of-the-art models, baseline methods are implemented for both the local grading and the biopsy-level scoring.

First, regarding the weakly-supervised patch-level Gleason grading, a model similar to \cite{Campanella2019Clinical-gradeImages} for MIL classification was implemented. This method, hereafter referred to as Global-Assignment, consists in assigning the global label to each patch of the WSI, and training a CNN using this pseudo-supervised dataset. In order to reduce noise on the pseudo-labels, only slides with one unique Gleason grade were used during training. Regarding the network employed, we trained a model using the same architecture, optimizer and learning rate schedule than we used in our student model. Furthermore, state-of-the-art methods for MIL aggregation were used in the teacher model to compare to the proposed max-pooling operation. Concretely, recent methods focus on using attention-based mechanisms for embedding-based aggregation in binary MIL classification tasks \cite{Ilse2018Attention-basedLearning}. The gated attention aggregation, referred to as AttMIL, was adapted to the instance-based multi-class aggregation use-case to obtain global predictions per each class $k$, $Y_k$, from the instance-level predictions, $y_{i,k}$ such that: $Y_{k} = \sum_{i}a_{i,k}y_{i,k}$. Thus, the attention weights, $a_{i,k}$, determine the contribution of each patch $i$ in the global prediction for each class via the features extracted by the CNN, 
$\mathbf{z}_i \in \mathbb{R}^{M}$, and the trainable parameters $\mathbf{V}\in \mathbb{R}^{L\times M}$, $\mathbf{U}\in \mathbb{R}^{L\times M}$ and $\mathbf{W}=[\mathbf{w}_0,...,\mathbf{w}_{K-1}]\in \mathbb{R}^{L\times K}$ as follows:


\begin{equation}
\label{eq:attMIL}
a_{i,k} = \frac{exp\{\mathbf{w}_k^{\top}(\tanh(\mathbf{V}\mathbf{z}_i)\odot sigm(\mathbf{U}\mathbf{z}_i))\}}{\sum_{i,k}exp\{\mathbf{w}_k^{\top}(\tanh(\mathbf{V}\mathbf{z}_i)\odot sigm(\mathbf{U}\mathbf{z}_i))\}}
\end{equation}

where $\tanh(\cdot)$ and $sigm(\cdot)$ are non-linearity functions, and $\odot$ an element-wise multiplication. The number of features extracted by the CNN is $M=512$ per instance, which are reduced to $L=128$ during the attention mechanism.

The global Gleason score and Grade Group was predicted also using previously proposed methods based on the percentage of each cancerous grade in the tissue (GG\%) using a kNN model as in \cite{Nagpal2019DevelopmentCancer} and a MLP used in \cite{Silva-rodriguez2020GoingDetection}. It is noteworthy to mention that other learnable aggregation functions were tested to obtain the embedding of instance-level features instead of the proposed average pooling. In particular, AttentionMIL \cite{Ilse2018Attention-basedLearning} and miGraph \cite{Zhou2009Multi-instanceSamples}\footnote{Based on the implementation proposed by Lammel et al. in: https://github.com/manuSrep/miGraphPy.} were evaluated. Nevertheless, these methods did not perform properly. Obtaining the Gleason score, by its very definition, involves obtaining the percentage of cancerous patterns in the biopsy, which does not match the formulation of the MIL methods (Equation \ref{eq:MIL}). It is noteworthy to mention that, in the weakly-supervised patch-level grading, we solve this limitation by using the presence of Gleason grades as global label, which fits the MIL formulation. We also performed an extensive comparison with previous results obtained in the same test subsets in \cite{Silva-rodriguez2020GoingDetection}. In that work, referred to as Supervised, a supervised CNN is trained using pixel-level annotations of Gleason grades performed by expert pathologists on WSIs from SICAP database. 

\subsection{Results}
\subsubsection{Grading of Local Patterns}
\label{ssec:exp2}

The figures of merit obtained using the teacher and student models on the different external datasets are presented in Table \ref{tab:resGleasonGrading}. In this table, we also report the results obtained in \cite{Silva-rodriguez2020GoingDetection}, who resort to supervised training, and those achieved by employing the the baseline approaches. Furthermore, we include the confusion matrices associated to the obtained results in Fig. \ref{fig:cm2}.

First, we will focus on the discussion of the results obtained by the different weakly supervised settings and the behavior of the teacher-student pair for this task. We can observe that teacher model achieved an inter-dataset average $\kappa$ of $0.69$ and $0.71$ using max and AttMIL as aggregation functions, respectively. This represents a significant improvement compared to the Global-Assignment model (average $\kappa = 0.47$), which is limited by the noise introduced in the patches labeled from cancerous biopsies using the global label. Although similar inter-dataset $\kappa$ is obtained for max and AttMIL aggregation functions in the teacher model, the former shows most promising results for Gleason grades differentiation. In particular, AttMIL achieves an inter-dataset average F1 of $0.6360$, $0.7140$ and $0.6601$ for GG3, GG4 and GG5, respectively. This shows the benefit of using attention mechanisms when training the teacher model, which enforces the model to focus on different patches to aggregate the instance-level predictions. Nevertheless, results shift when training the student model using the teacher model's predictions as pseudo-labels. We can observe that the student model using max aggregation obtained and inter-dataset average $\kappa$ of $0.82$ and a F1 of $0.77$, an improvement of $8\%$ and $12\%$, respectively, compared to its corresponding teacher model. On the other hand, the student model does not show any improvement when using AttMIL as aggregation function. In the interpretation of these results, the process of label refinement between the teacher and the student models (see Section \ref{ssec:meth2}) plays a fundamental role. In this process, false negative patches from positive bags are discarded during training, while false positive instances cannot be detected. These instances classified wrongly as cancerous by the teacher model are the main source of noise in the student model training. Furthermore, we observe that max-pooling aggregation results on an inter-dataset precision in the detection of cancerous instances of $0.95$, whereas AttMIL aggregation obtains only $0.26$. This difference can be explained by the fact that the slide-level max pooling operation in the teacher model architecture produces backpropagation of the weights for only cancerous patterns that the model classifies with high confidence. Although this phenomenon increases the number of false negative for cancerous classes, these samples are discarded during the aforementioned label-refinement process. Thus, our proposed framework using max-pooling as MIL aggregation function in the teacher model does not introduce noise during the pseudo-labeling process, which results on a better performance of the student model.

Regarding previous state-of-the-art methods for patch-level Gleason grading based on supervised training on pixel-level annotations, our proposed teacher-student model using max-aggregation compared also favorable. In the supervised method in \cite{Silva-rodriguez2020GoingDetection}, which employs a CNN trained on annotated patches from SICAP database, a consistent drop in the $\kappa$ metric was observed across the three datasets: SICAP test subset ($\kappa = 0.77$) to ARVANITI ($\kappa = 0.64$) and GERTYCH ($\kappa = 0.51$). In contrast, our weakly-supervised model obtained similar results in the three external datasets, with $\kappa = 0.83$ on SICAP, $\kappa = 0.79$ on ARVANITI and $\kappa = 0.82$ on GERTYCH. These values demonstrate that the proposed weakly supervised pipeline outperforms the methodology presented in \cite{Silva-rodriguez2020GoingDetection}, showing a higher generalization ability and requiring a weaker supervision during training. This comparison extends to other prior works using supervised learning, which reach $k$ values of $0.55/0.49$ in \cite{Arvaniti2018AutomatedLearning} and $0.61$ in \cite{Nir2018AutomaticExperts} under different setups. The reason for superior performance of the proposed weakly-supervised strategy could be due to the bias of the annotator produced in the supervised learning scenario, which is not present when using global-labels in our pipeline. Furthermore, the difficulty of obtaining large annotated datasets with heterogeneous patterns can also reduce the performance of fully supervised learning approaches. These benefits over previously proposed methods in the literature outweigh the disadvantages of the proposed strategy. In particular, we might identify as potential limitations the large computational requirements of processing all the patches of a biopsy in each iteration during the training of the teacher model, and the need to use large datasets for correct generalization. These drawbacks, however, are an inherent characteristic of weakly supervised strategies.

\begin{figure}[htb]
\captionsetup[subfloat]{farskip=1pt,captionskip=0.8pt}
    \centering
      \subfloat[\label{fig1a}]{\includegraphics[width=.235\linewidth, frame]{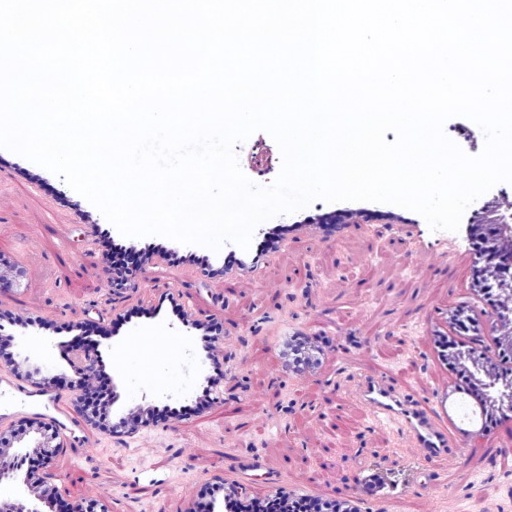}}
      \hspace*{\fill}
      \subfloat[\label{fig1b}]{\includegraphics[width=.235\linewidth, frame]{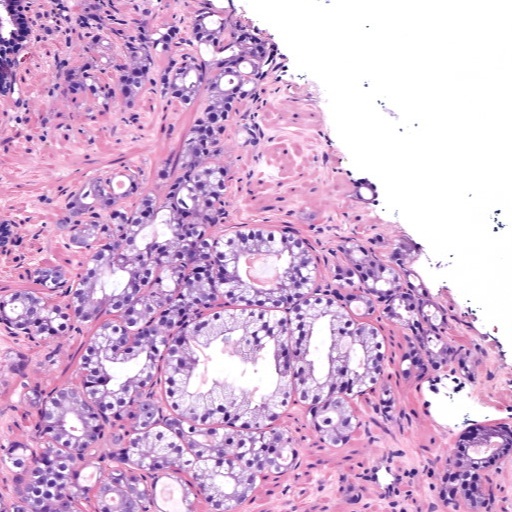}}
      \hspace*{\fill}
      \subfloat[\label{fig1c}]{\includegraphics[width=.235\linewidth, frame]{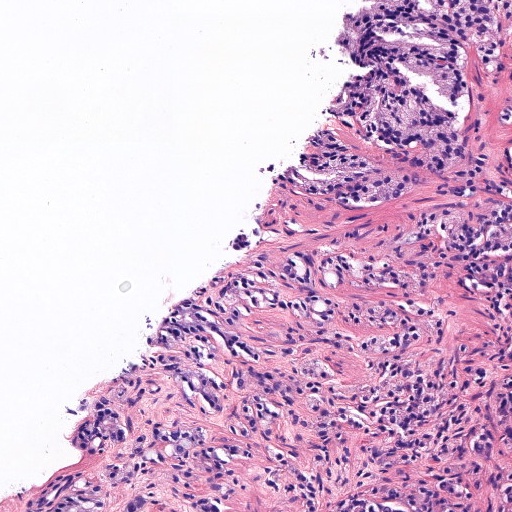}}
      \hspace*{\fill}
      \subfloat[\label{fig1d}]{\includegraphics[width=.235\linewidth, frame]{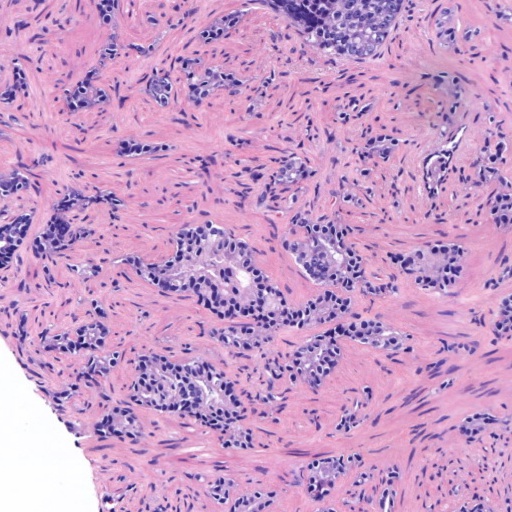}}
    \caption{Histology regions with mixed Gleason grades from SICAP dataset. (a): region containing benign and GG3 glands, (b): region containing GG3 and GG4 glandular structures, (c) and (d): region containing GG4 and GG5 patterns. GG: Gleason grade.}
    \label{fig10}
\end{figure}

\begin{figure*}[htb]
\captionsetup[subfloat]{farskip=1pt,captionskip=0.8pt}
    \begin{center}
    
      \subfloat[\label{fig:cm2a}]{\includegraphics[width=.32\linewidth]{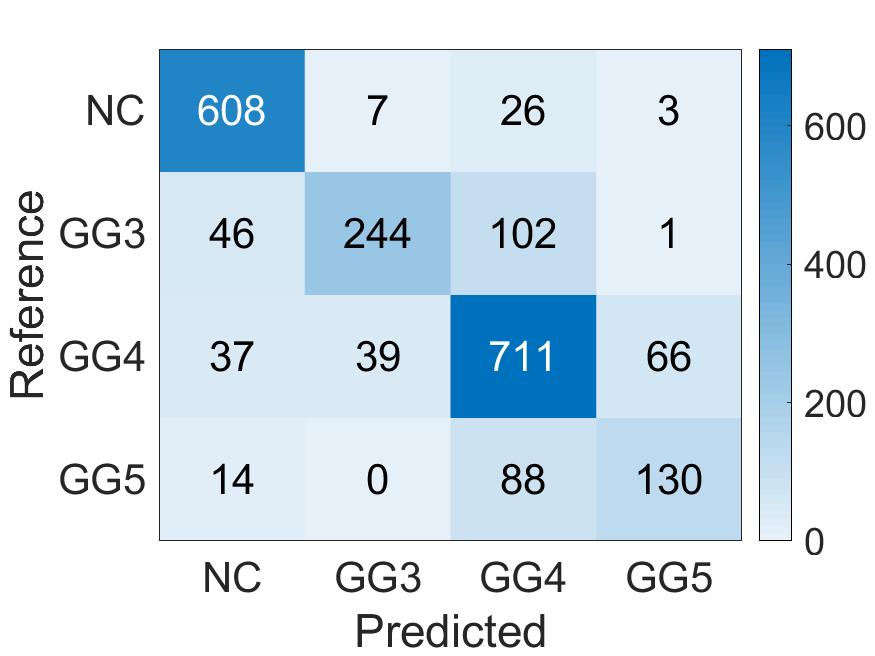}}
      \hspace*{\fill}
      \subfloat[\label{fig:cm2b}]{\includegraphics[width=.32\linewidth]{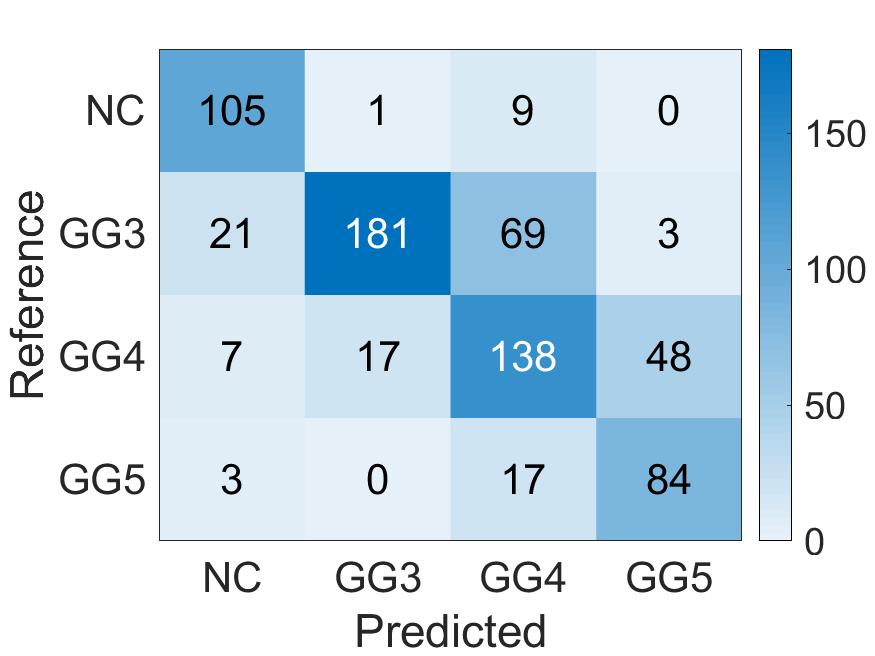}}
      \hspace*{\fill}
      \subfloat[\label{fig:cm2c}]{\includegraphics[width=.32\linewidth]{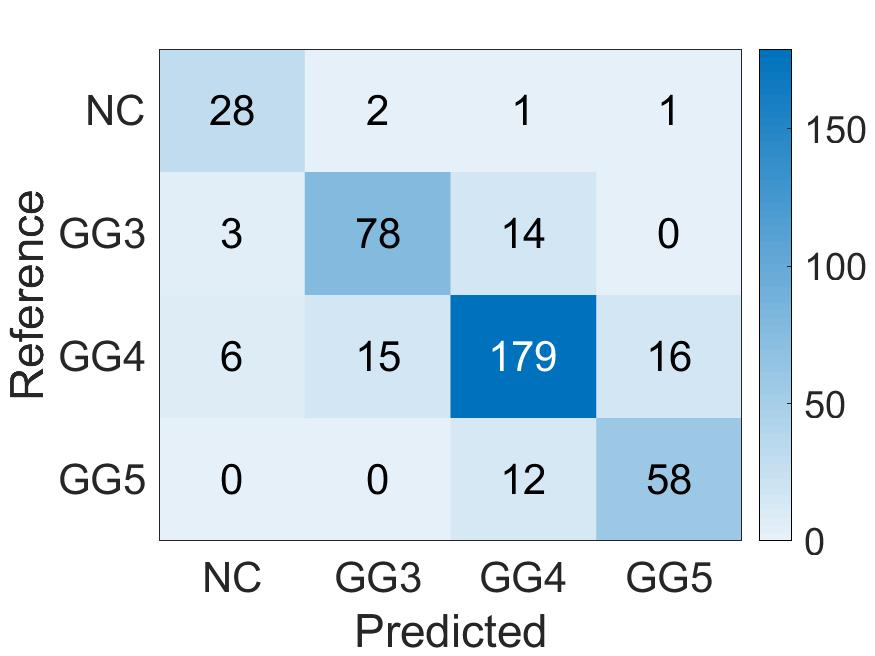}}
      \hspace*{\fill}
      
    \caption{Confusion Matrix of the patch-level Gleason grades prediction done by Student CNN on the different test cohorts. (a): SICAP; (b): ARVANITI; (c): GERTYCH.}
    \label{fig:cm2}
    \end{center}
\end{figure*}

Finally, we would like to highlight the limitations observed to evaluate patch-level Gleason grading models in different, heterogeneous datasets. Although similar $\kappa$ values were obtained across the three external datasets, differences can be observed when focusing on concrete figures of merit. For instance, the best results are obtained on GERTYCH dataset (average F1 of $0.82$), whereas the worse results are reported on ARVANITI dataset (average F1 of $0.73$). Although the overall results are similar on SICAP dataset (average F1 of $0.75$), the student model performs poorly on the GG5 class (F1 of $0.60$) and gives the best results on the NC class (F1 of $0.90$). These differences could be related to different reasons. For instance, in each dataset the balance of the classes is not equal (see Table \ref{tab:dataPatches}). Precisely, SICAP dataset presents a larger proportion of NC patches, while the proportion of GG5 cases is lower. In addition, SICAP dataset contains patches with mixed Gleason grades, whose label is assigned by majority voting. In these cases, the CNN could be mixing the features of the different classes, thus hampering the obtained performance. Examples of these patches are presented in Figure \ref{fig10}. Another limitation for Gleason grading assessment is the well-known inter-pathologist variability. Thus, specific patterns could be annotated with different Gleason grades by pathologists. This variability was quantified at patch level by Arvaniti et al. \cite{Arvaniti2018AutomatedLearning}, obtaining a $\kappa$ of $0.65$. This fact enhances the importance of testing the proposed methods across different datasets to ensure the generalization capability of the CAD systems for Gleason grading.

\begin{table*}[htb]

    \centering
    \caption{Results for the patch-level Gleason grades classification performed by the different approaches on the different test cohorts. The metrics presented are the accuracy (ACC), the F1-Score (F1) and the Cohen's quadratic kappa ($\kappa$). Furthermore, precision and sensitivity are indicated for the non-cancerous class. Bold numbers highlight the best performing method. NC: non cancerous, GG: Gleason grade.}
    \label{tab:resGleasonGrading}

    \begin{tabular}{|l|c|ccccc|c|c|c|}
    
    \hline
    \multicolumn{1}{|c|}{\multirow{2}{*}{\textbf{Method}}} & \multicolumn{7}{c|}{\textbf{Grading}} & \multicolumn{2}{c|}{\textbf{Binary C/NC}} \\ \cline{2-10}
     & \textbf{ACC} & \multicolumn{5}{c|}{\textbf{F1}} & \textbf{$\kappa$} & \textbf{Sensitivity} & \textbf{Precision} \\ 
     & & NC & GG3 & GG4 & GG5 & Avg. & & & \\
    \hline
    \hline
    \multicolumn{10}{c}{Other settings} \\
    \hline
    \hline
   Arvaniti et al. \cite{Arvaniti2018AutomatedLearning} ($2018$)* & $-$ & $-$ & $-$ & $-$ & $-$ &  $-$ & $0.55/0.49$ & $-$ & $-$ \\
    \hline
    Nir et al. \cite{Nir2018AutomaticExperts} ($2018$)** & $-$ & $-$ & $-$ & $-$ & $-$ & $-$ & $0.60$ & $-$ & $-$ \\
    \hline
    \hline
    \multicolumn{10}{c}{SICAP} \\
    \hline
    \hline
    Supervised \cite{Silva-rodriguez2020GoingDetection} ($2020$) & $0.67$ & $0.86$ & $0.59$ & $0.54$ & $0.61$ & $0.65$ & $0.77$  & - & -\\
    \hline    
    Global Assignment & $0.5056$  & $0.0756$  & $0.4408$  & $0.7441$  & $0.0952$ & $0.3389$ & $0.4659$  & $0.7321$ & $0.0395$ \\
    \hline
    Teacher - Max & $0.7229$ & $0.7884$ & $0.6426$ & $0.6426$ & $0.2173$ & $0.6047$ & $0.6360$  & $0.6631$ & $\mathbf{0.9720}$ \\
    \hline
    Teacher - AttMIL & $0.6555$ & $0.6573$ & $0.5441$ & $0.7680$ & $0.4834$ & $0.6132$ & $0.7251$ & $0.9118$ & $0.5139$ \\
    \hline
    Student - Max & $\mathbf{0.7978}$ & $\mathbf{0.9014}$ & $\mathbf{0.7144}$ & $\mathbf{0.7988}$ & $\mathbf{0.6018}$ & $\mathbf{0.7541}$ & $\mathbf{0.8303}$  & $0.8624$ & $0.9441$ \\
    \hline
    Student - AttMIL & $0.6639$ & $0.6538$ & $0.5639$ & $0.7605$ & $0.5446$ & $0.6307$ & $0.7287$ & $\mathbf{0.9389}$ & $0.5015$ \\
    \hline
    \hline
    \multicolumn{10}{c}{ARVANITI} \\
    \hline
    \hline
    Supervised \cite{Silva-rodriguez2020GoingDetection} ($2020$) & $0.5861$ & $0.566$ & $0.6858$ & $0.4699$ & $0.5603$ & $0.5702$ & $0.641$  & - & -\\
    \hline    
    Global Assignment & $0.5547$  & $0.0172$  & $0.6747$  & $0.6120$  & $0.2051$ & $0.3772$ & $0.5012$  & $0.6442$ & $0.0086$ \\
    \hline
    Teacher - Max & $0.7055$ & $0.7266$ & $0.7308$ & $\mathbf{0.6827}$ & $0.6666$ & $0.7017$ & $0.7567$  & $0.5892$ & $\mathbf{0.9478}$ \\
    \hline
    Teacher - AttMIL & $0.6557$ & $0.2714$ & $0.7258$ & $0.6470$ & $0.7253$ & $0.5924$ & $0.7162$ & $0.7600$ & $0.1652$ \\
    \hline
    Student - Max & $\mathbf{0.7226}$ & $\mathbf{0.8366}$ & $\mathbf{0.7653}$ & $0.6230$ & $\mathbf{0.7029}$ & $\mathbf{0.7319}$ & $\mathbf{0.7930}$  & $\mathbf{0.7721}$ & $0.9130$ \\
    \hline
    Student - AttMIL & $0.6358$ & $0.1269$ & $0.7129$ & $0.6266$ & $0.7333$ & $0.5499$ & $0.6891$ & $0.7272$ & $0.0695$ \\
    \hline
    \hline
    \multicolumn{10}{c}{GERTYCH} \\
    \hline
    \hline
    Supervised \cite{Silva-rodriguez2020GoingDetection} ($2020$) & $0.5136$ & $0.2901$ & $0.6162$ & $0.499$ & $0.4959$ & $0.4753$ & $0.5116$  & - & -\\
    \hline    
    Global Assignment & $0.5622$  & $0.0143$  & $0.6932$  & $0.7615$  & $0.2672$ & $0.4340$ & $0.5312$  & $0.4623$ & $0.0073$ \\
    \hline
    Teacher - Max & $0.7893$ & $0.6976$ & $0.7958$ & $0.8356$ & $0.6666$ & $0.7489$ & $0.6943$  & $0.5556$ & $\mathbf{0.9375}$ \\
    \hline
    Teacher - AttMIL & $0.6803$ & $0.1621$ & $0.6394$ & $0.7277$ & $0.7716$ & $0.5752$ & $0.6934$ & $0.6000$ & $0.0937$ \\
    \hline
    Student - Max & $\mathbf{0.8305}$ & $\mathbf{0.8115}$ & $\mathbf{0.8210}$ & $\mathbf{0.8483}$ & $0.8000$ & $\mathbf{0.8202}$ & $\mathbf{0.8264}$  & $\mathbf{0.7568}$ & $0.8750$ \\
    \hline
    Student - AttMIL & $0.7070$ & $0.1111$ & $0.6367$ & $0.7609$ & $\mathbf{0.8507}$ & $0.5898$ & $0.7313$ & $0.5000$ & $0.0625$ \\
    \hline
     \multicolumn{8}{l}{\scriptsize{* Results reported on different patch size and resolutions}.}\\
     \multicolumn{8}{l}{\scriptsize{** Results reported on a different (private) dataset}.}
    \end{tabular}

\end{table*}

\subsubsection{Biopsy-Level Scoring}
\label{ssec:exp3}

\begin{figure}[htb]
\captionsetup[subfloat]{farskip=1pt,captionskip=0.8pt}
    \begin{center}
    
      \subfloat[\label{fig:cm3a}]{\includegraphics[width=.45\linewidth]{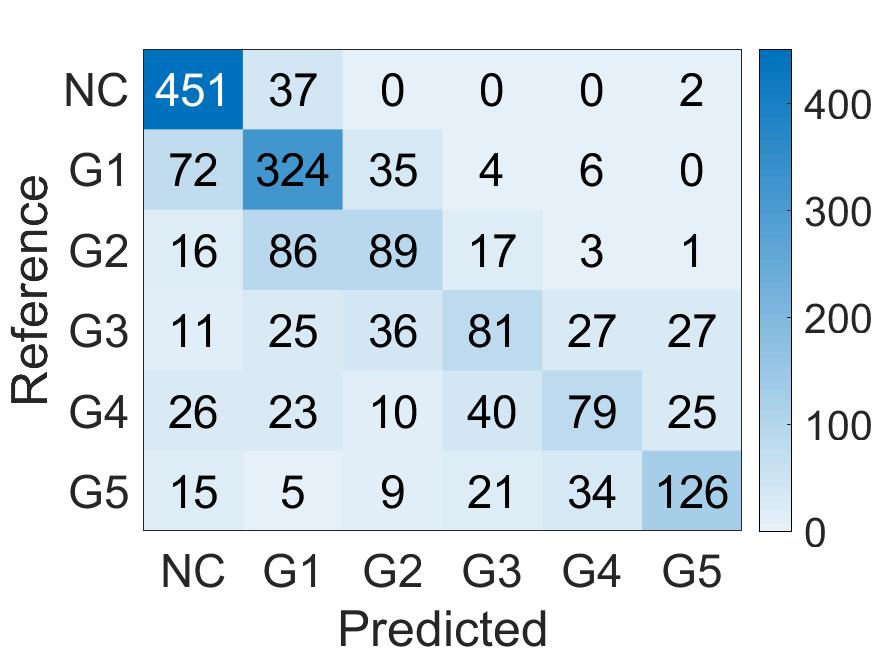}}
      \hspace*{\fill}
      \subfloat[\label{fig:cm3b}]{\includegraphics[width=.45\linewidth]{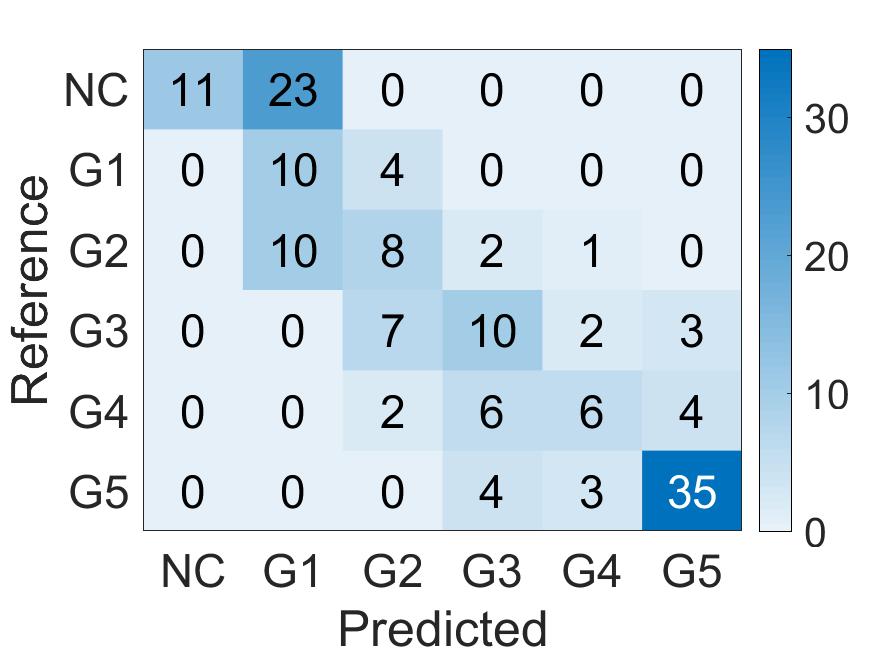}}
      \hspace*{\fill}
      
    \caption{Confusion Matrix of the biopsy-level Grade Group prediction done by Student CNN features and k-Nearest Neighbor classifier, on the two test cohorts. (a): PANDA; (b): SICAP}
    \label{fig:cm3}
    \end{center}
\end{figure}

Table \ref{tab:resGleasonScoring} reports the results obtained by the proposed approaches based on the student features, as well as those from the baseline methods (GG\%). Also, results reported in previous works are indicated. Similarly, the confusion matrices of the Grade Group predictions using Student model as feature extractor are illustrated in Fig. \ref{fig:cm3}.

\begin{table}[htb]

    \centering
    \caption{Results of the biopsy-level Gleason scoring in the test subsets. The metric presented is the Cohen's quadratic kappa ($\kappa$).}
    \label{tab:resGleasonScoring}

    \begin{tabular}{|l|c|c|}

    \hline
    \multicolumn{1}{|c|}{\textbf{Method}} & \textbf{Gleason Score} & \textbf{Grade Group}\\
    \hline
    \hline
    \multicolumn{3}{c}{Other settings} \\
    \hline
    \hline
    Arvaniti et al. \cite{Arvaniti2018AutomatedLearning} ($2018$) & $0.75/0.71$  & $-$ \\
    \hline
    Bulten et al. \cite{Bulten2020AutomatedStudy} ($2020$) & $-$  & $0.85$*$/0.72$** \\
    \hline
    Strom et al. \cite{Strom2020ArtificialStudy} ($2020$) & $-$  & $0.91$*$/0.82$** \\
    \hline
    Otalora et al. \cite{Otalora2020AGrading} ($2020$) & $-$  & $0.44$*** \\
    \hline
    \hline
    \multicolumn{3}{c}{PANDA} \\
    \hline
    \hline
    GG\% + kNN & $0.7936$ & $0.8152$ \\
    \hline
    GG\% + MLP & $\mathbf{0.8054}$   & $0.8229$  \\
    \hline
    Features + Average + kNN & $0.7773$ &
    $0.7927$  \\
    \hline
    Features + Average + MLP & $0.7954$  & $\mathbf{0.8245}$ \\
    \hline
    \hline
    \multicolumn{3}{c}{SICAP} \\
    \hline
    \hline
    Supervised \cite{Silva-rodriguez2020GoingDetection} & $0.8177$ & $-$ \\ 
    \hline
    GG\% + kNN & $0.5942$ & $0.5221$  \\
    \hline
    GG\% + MLP &  $0.4861$ & $0.5082$ \\
    \hline
    Features + Average + kNN & $\mathbf{0.8299}$ &
    $\mathbf{0.8854}$  \\
    \hline
    Features + Average + MLP & $0.3847$  & $0.4421$ \\
    \hline
    \multicolumn{3}{l}{\scriptsize{* Results on test subset}.}\\
    \multicolumn{3}{l}{\scriptsize{** Results reported on external datasets}.}\\
    \multicolumn{3}{l}{\scriptsize{*** The used dataset does not include benign biopsies}.}
    \end{tabular}

\end{table}

Our proposed model based on aggregating patch-level features and kNN classifier outperformed previously state-of-the-art methods based on the use of the percentage of each Gleason grade in the tissue, reaching an average $\kappa$ of $0.84$ for both datasets. Although the results were similar for all models when testing on biopsies from the same center as in the training cohort (PANDA), the performance of neural-networks-based methods dropped on the external dataset (SICAP). The classification stage on neural networks showed overfitting to the training set characteristics in both Gleason grade percentage calculation (based on patch-level classification) and in the global score prediction using the Student model features. The non-parametric use of raw features and kNN generalized better on external datasets. The obtained results are promising, being most of the errors between adjacent classes. Moreover, the model differentiated well critical cases such Grade Group $2$ and $3$, whose main difference is the balance between Gleason grade $3$ and $4$ in the tissue (see Fig. \ref{fig:cm3}). It is noteworthy to mention that the different results reported from other works are evaluated under different datasets and training conditions. Thus, direct comparison to those works is unfair.

The obtained results are in line with previous literature for global Gleason scoring. The proposed method is comparable against works that use strong supervision via pixel-level annotations, i.e., Arvaniti et al. \cite{Arvaniti2018AutomatedLearning} ($\kappa = 0.75$ for Gleason scoring) and Bulten at al. \cite{Bulten2020AutomatedStudy} ($\kappa = 0.85$ for Grade Group scoring), as well as works that use only global biopsy-level labels, i.e., Strom et al. \cite{Strom2020ArtificialStudy} ($\kappa = 0.91$ for Grade Group scoring) and Otálora et al. \cite{Otalora2020AGrading} ($\kappa = 0.44$ for Grade Group scoring). In accordance with the observations in our work, methods based on the Gleason grades proportion in the tissue suffer a performance drop on external datasets ($\kappa = 0.72$ in Bulten et al. and $\kappa = 0.82$ in Strom et al.) Finally, we would like to highlight the difficulty of establishing comparisons among different datasets, since most of the presented results are at the level of inter-pathologist variability for Gleason scoring. Different works have quantified the inter-observer variability on kappa values of 0.71 by Arvaniti et al. \cite{Arvaniti2018AutomatedLearning} or ranging 0.726-0.869 by Bulten et al. \cite{Bulten2020AutomatedStudy}.

\begin{figure*}
    \centering
    
    \renewcommand{\thesubfigure}{a}
    \subfloat[\label{regiona}]{\resizebox{1\textwidth}{!}{
    \begin{tikzpicture}[spy using outlines={circle,yellow,magnification=2,size=3cm, connect spies}]
    \node {\pgfimage[interpolate=true,height=5cm]{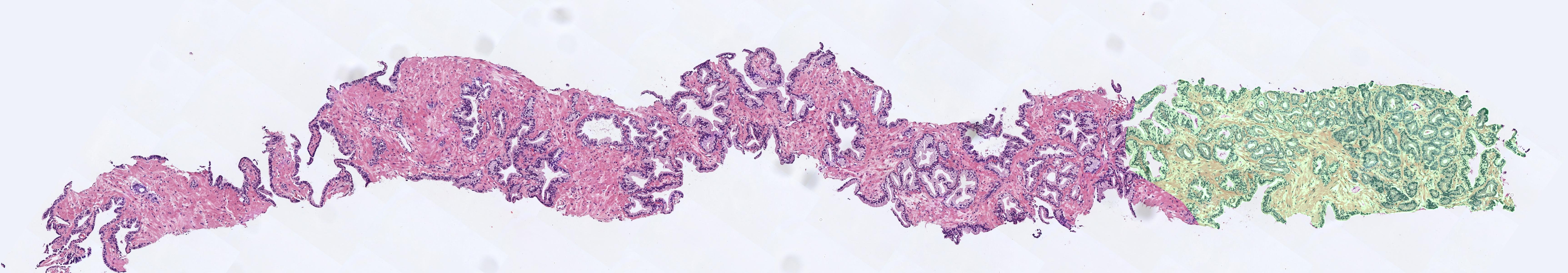}};
    \spy on (11.5,0.3) in node [left] at (9.5, 2.5);
    \end{tikzpicture}}}

    \renewcommand{\thesubfigure}{b}
    \subfloat[\label{regionb}]{\resizebox{1\textwidth}{!}{
    \begin{tikzpicture}[spy using outlines={circle,yellow,magnification=4,size=3cm, connect spies}]
    \node {\pgfimage[interpolate=true,height=5cm]{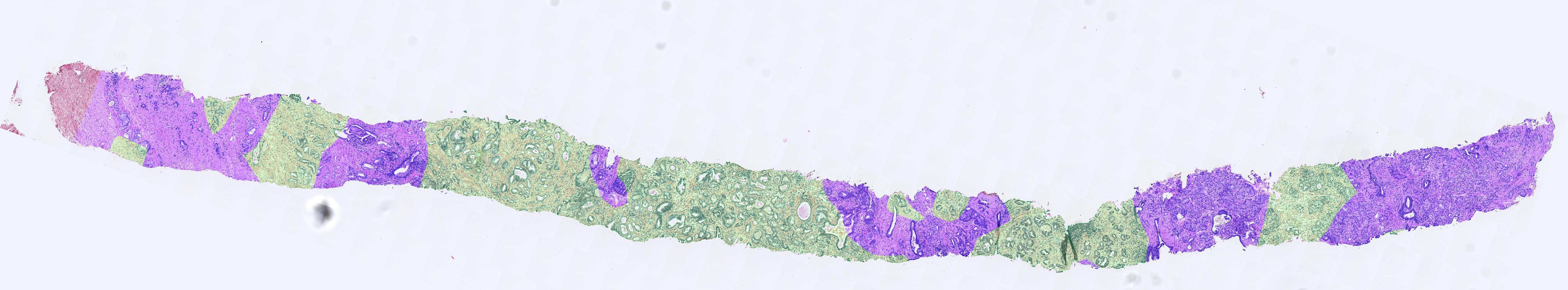}};
    \spy on (-0.4,-1.4) in node [left] at (6,1);
    \spy on (11.5,-1) in node [left] at (10,1);
    \spy on (-5,0) in node [left] at (2,1);
    \end{tikzpicture}}}

    \renewcommand{\thesubfigure}{c}
    \subfloat[\label{regionc}]{\resizebox{1\textwidth}{!}{
    \begin{tikzpicture}[spy using outlines={circle,yellow,magnification=2,size=2cm, connect spies}]
    \node {\pgfimage[interpolate=true,height=5cm]{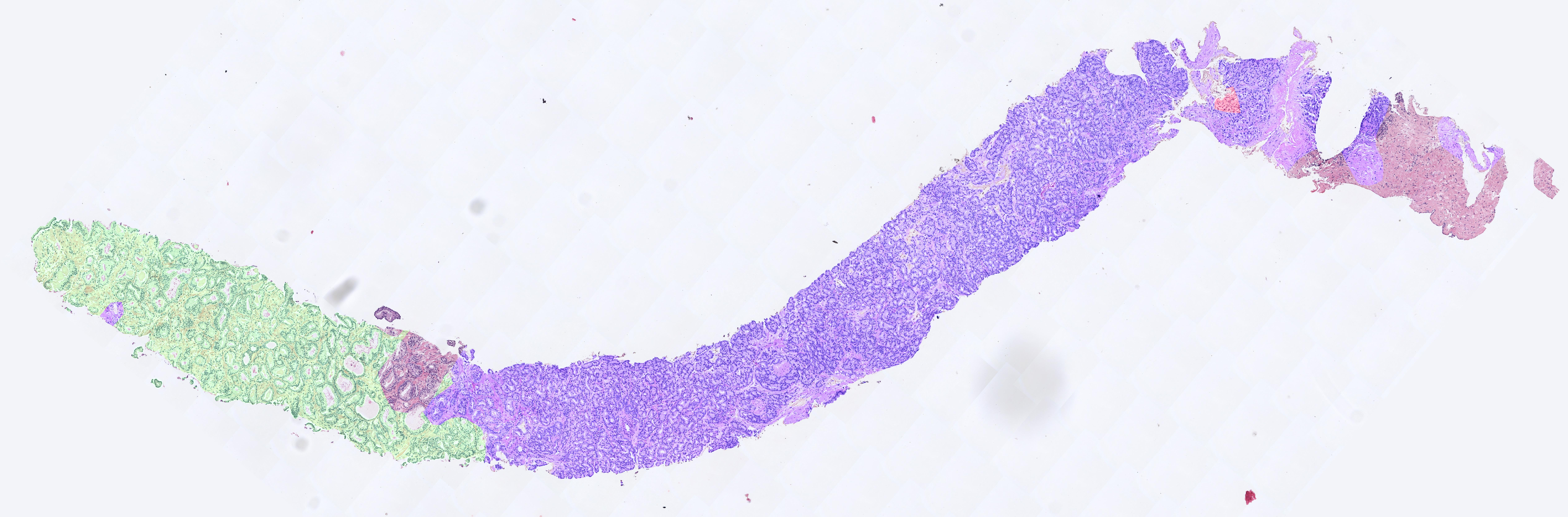}};
    \spy on (-5,-1) in node [left] at (-5,1);
    \spy on (-1,-1.5) in node [left] at (-2.5,1);
    \spy on (2,0.5) in node [left] at (0,1);
    \spy on (4.5,1.5) in node [left] at (6,-1.5);
    \end{tikzpicture}}}

    \renewcommand{\thesubfigure}{d}
    \subfloat[\label{regiond}]{\resizebox{1\textwidth}{!}{
    \begin{tikzpicture}[spy using outlines={circle,yellow,magnification=8,size=3cm, connect spies}]
    \node {\pgfimage[interpolate=true,height=5cm]{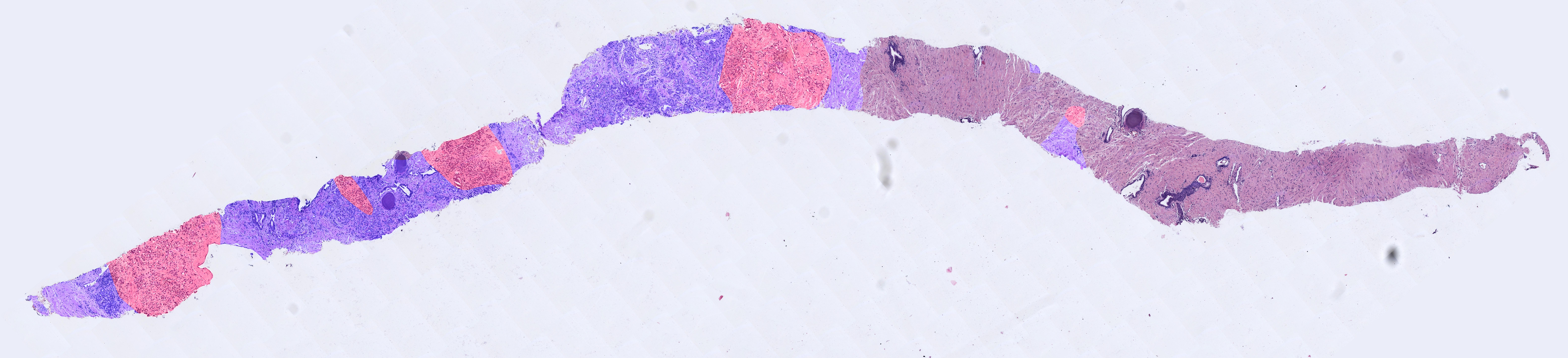}};
    \spy on (-8.5,-1.5) in node [left] at (-6,1.5);
    \spy on (-1.9,1.25) in node [left] at (0,-1.3);
    \spy on (-0.1,1.5) in node [left] at (4,-1.3);
    \end{tikzpicture}}}

    \caption{Examples of the proposed CAD system based on Self-Learning CNNs performance on the external SICAP dataset. In green: Gleason grade $3$; blue: Gleason grade $4$ and red: Gleason grade $5$. The reference and predicted Grade Group (Reference - Predicted) are: (a): ($2$ - $1$); (b): ($2$ - $2$); (c): ($3$ - $3$); (d): ($5$ - $5$).}
    \label{fig:slides}
\end{figure*}

Representative examples of the obtained results using the CAD system on the external SICAP dataset are presented in Fig. \ref{fig:slides}. The pixel-level heatmaps are obtained by bilinearly interpolating the patch-level predicted probabilities of the closest patches in terms of euclidean distance. Then, the class with highest probability is assigned to each pixel. This figure is organized as follows: each row constitutes a case, and the Gleason grades in the tissue are highlighted in different colors. Also, the Grade Group predicted and the ground truth are indicated.

\section{Conclusions}
\label{sec:conclusions}

In this work, we have proposed a novel self-learning CNN strategy to perform both Gleason grading of local cancerous patterns and global scoring of prostate WSIs using only the global Gleason score during training. Our proposed framework is composed by a novel teacher model based on max-pooling of patch-level inferences of Gleason grades able to perform local classification of Gleason grade using biopsy-level labels. Based on the output of the teacher model and a label refinement post-processing, we propose the training of a patch-level student model on a pseudo-supervised dataset. In the experimental stage, we validate the patch-level classification on three different external datasets. The student model reaches an inter-datasets average Cohen's quadratic kappa ($\kappa$) of $0.82$ and an f1-score of $0.77$. Our results outperformed previous works based on supervised learning with pixel-level annotations. Moreover, the results between the different test cohorts were similar, while previous supervised methods experimented a drop in performance when testing on external test images. Our proposed weakly-supervised method generalizes better than supervised methods for local Gleason grading, due to the absence of annotator bias and the capability of being trained on large heterogeneous datasets.  

Then, the features learned by the patch-level trained models were used to predict the global Grade Group via an average aggregation and a linear classification layer. The method was tested on two different datasets, reaching an average $\kappa$ of $0.84$. This method was compared with the main approach in the literature for Grade Group prediction using the percentage of the different Gleason grades in the tissue. Our feature-based model showed to better generalize pathologist scoring biopsies than previous approaches.

The promising results presented in this work represent a significant advance in the literature of prostate histology. Using weakly-supervised learning it is possible to grade local patterns in gigapixel WSIs outperforming supervised methods which require laborious annotations by expert pathologists. Further research will focus on studying and improving the image-normalization process of prostate histology samples to use CAD systems in external datasets, which is a vital step for a successful generalization. Also, the proposed weakly-supervised models will be refined to decrease the number of biopsies required during training.




\bibliography{references_manual}
\bibliographystyle{IEEtran}

\end{document}